\magnification=1200
\baselineskip=18truept
\input epsf

\def\preprint{Y}
\def\draftversion{N}
\def\cap{\hsize=4.5in}

\if \draftversion Y


\fi

\def\figure#1#2#3{\if \preprint Y \midinsert \epsfxsize=#3truein
\centerline{\epsffile{figure_#1_eps}} \halign{##\hfill\quad
&\vtop{\parindent=0pt \hsize=5.5in \strut## \strut}\cr {\bf Figure
#1}&#2 \cr} \endinsert \fi}

\def\figureb#1#2{\if \preprint N \midinsert \epsfxsize=#2truein
\centerline{\epsffile{figure_#1_eps}} \halign{##\hfill\quad
&\vtop{\parindent=0pt \hsize=5.5in \strut## \strut}\cr \cr \cr
\cr \cr \cr  {\bf Figure #1} \cr} \endinsert \fi}

\def\captionone{\cap 
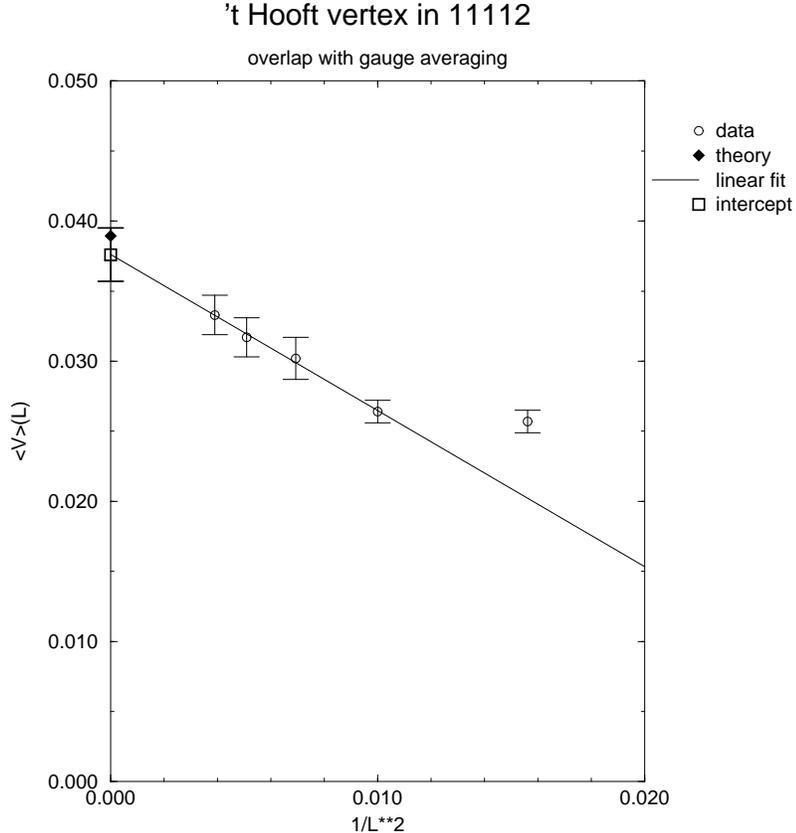
Data for $L=8,10,12,14,16$ versus $1/L^2 \propto a^2$, 
a linear
fit to the points $L\ge 10$, the continuum result (rhombus at $a=0$),
and our estimate for the continuum result from the data (square
with error bar at $a=0$).}

\def\chiral{ {\hat\theta} }

\line{\hfill RU-97-01}
\line{\hfill UW/PT-97-01}
\line{\hfill DOE/ER-40561-310-INT97-00-161}
\vskip 2truecm
\centerline{\bf Finite size corrections in two dimensional gauge theories}
\centerline{\bf and a quantitative chiral test of the overlap.}

\vskip 1truecm
\centerline{Yoshio Kikukawa${}^{a}$\footnote{*}{Permanent Address: 
Department of Physics, Kyoto University, Kyoto 606-01, Japan.
}, Rajamani Narayanan${}^{b}$
and Herbert Neuberger${}^{a}$}
\vskip .5truecm

\centerline {${}^a$ Department of Physics and Astronomy}
\centerline {Rutgers University, Piscataway, NJ 08855-0849}
\centerline {${}^b$ Institute for Nuclear Theory, Box 351550}
\centerline {University of Washington, Seattle, WA 98195-1550}
\vskip 1.5truecm

\centerline{\bf Abstract}
\vskip 0.75truecm
An argument is presented for a certain universality of finite
size corrections in two dimensional gauge theories. In the abelian case
a direct calculation is carried out for a particular chiral model. 
The analytical
result confirms the above universality and that the 
't Hooft vertex previously measured using the overlap 
smoothly approaches the correct continuum limit
within statistical errors.

\vfill
\eject
                                                       
This note contains two main parts: In the 
first we apply some analytical methods
to the problem of finite size effects in two dimensional gauge theories.
In the second part the theoretical results are exploited to interpret some
numerical data.

\noindent
{\bf 1. } Quite recently it was argued that two dimensional gauge theories 
exhibit a dynamical decoupling of the Hilbert space into a conformal 
theory and a massive sector [1]. 
The global ``flavor'' symmetries of the original model get elevated
to Kac--Moody algebras represented within the massless sector. 
Moreover, different theories can have
identical massive sectors, the latter being sensitive to 
only some very general
properties of the model. 

Here our concern with two dimensional gauge theories is 
limited to their
usability as benchmark cases against which proposals to 
regulate a chiral
gauge theory in any dimension should be tested. Such numerical
tests are carried out in a finite Euclidean volume, typically,
a torus. It is therefore useful to obtain as much exact 
information about
finite size effects in the continuum as possible.\footnote{*}
{This may have some applications to finite temperature problems.}

Although the arguments of Kutasov and Schwimmer apply more
directly at infinite volume, it is plausible that with judicious 
choices of
boundary conditions, their conclusions hold exactly also for 
selected theories
defined on finite tori. Assume we are in such a situation and that we
are computing the expectation value of some one-point observable. 
The observable
can be factorized into an operator acting within the massless 
sector and another acting within
the massive sector. Attaching the right power of the gauge coupling constant
the massless factor can be made dimensionless, so its expectation
value is a size independent pure number because of conformal invariance.
Therefore, any finite size correction must come from the massive factor.
Then, the Kutasov--Schwimmer universality extends to the finite
size correction, meaning that it can be computed in any
of a class of theories. In particular, one can always find a vector
representative of the massive sector. We conclude that finite size
corrections for any chiral model can be evaluated by looking at
appropriate finite size corrections in an associated vector model. 

The above holds equally in the abelian and non-abelian case. 
Actually, in
the abelian case the decoupling and the finite size effects are easier to
analyze. In previous work [2,3] we concentrated on a particular abelian
model called the 11112 model for its fermion content. 
The action of the 11112 model in Euclidean space is:
$$
S={1\over 4e_0^2} \int d^2 x F^2_{\mu\nu}
-\sum_{k=1}^4\int d^2 x \bar\chi_k \sigma_\mu 
(\partial_\mu+iA_\mu )\chi_k
-\int d^2 x \bar\psi \sigma^*_\mu (\partial_\mu+2iA_\mu )\psi ,\eqno{(1)}$$
where $\sigma_1=1$, $\sigma_2=i$ and $\mu =1,2$. 
In addition to this, we have to specify the 
boundary conditions on the fermions.
We can always choose one of the fermions to obey periodic boundary conditions
by a suitable redefinition of the gauge field. We will assume this is done
and choose $\psi$ to obey periodic boundary conditions. 
The four $q=1$ fermions obey the following boundary conditions:
$$\chi_f(x_\mu+ l \hat\mu)= e^{2\pi i b^f_\mu}\chi_f(x_\mu) .\eqno{(2)}$$
We restrict the $b^f_1$ to the interval $[-1/2 , 1/2 )$ and the $b^f_2$
to the interval $(-1/2 ,  1/2]$. Physically,
$b^f_\mu$ differing by integers mean exactly the same thing, but
in all subsequent formulae the symbols $b^f_\mu$ are assumed to reside
within the above ranges.

Due to the abelian $U(1)$ anomaly and instantons the dimensionless operator
$$V(x)={{\pi^2}\over
{e_0^4}} \chi_1(x) \chi_2(x) \chi_3(x) \chi_4(x)
\bar\psi(x) (\sigma\cdot\partial )\bar\psi(x),
\eqno{(3)}$$
which is the 't Hooft vertex in this model, 
gets a nonzero expectation
value.
Assuming clustering at infinite volume we evaluated this expectation value
in [2], obtaining 
$$
\langle V \rangle ={{e^{4\gamma}}\over {4\pi^3}}\approx 0.081\eqno{(4)}$$

In this case the factorization of $V$ could be seen explicitly quite easily in
a formal bosonized operator solution at infinite volume. To get a rigorous
formula in a finite volume one would need to worry about boundary conditions
and topological effects 
in the operator formalism and this we have not done.
Our objective here is 
to show by direct computation of the path integral
that the finite size corrections to 
$\langle V\rangle_l$ 
measured on a torus of physical size $l \times l$ indeed are universal
in the sense explained above. In particular we wish to see that
these finite size effects are identical to the
ones in a four flavor vector Schwinger model, which happens to be the
simplest associated model. We should emphasize that the
pertinent 't Hooft vertex operators are quite different in the two models:
In the chiral case we have six fermions and a derivative while in the vector
case we have eight fermions. In the chiral case
the operator does not commute with fermion number but
in the vector case it does. 

A detailed understanding of finite size corrections in the vector
case would lead one to guess that the above universality holds even without 
employing the Kutasov--Schwimmer argument directly. 
Thus, in
our previous work we already used the four flavor Schwinger model as
a source for an estimate of the finite size effects in the chiral case.
Here we shall present explicit proof and validate our previous
procedure. Moreover, by adopting the Kutasov--Schwimmer logic it
appears that a similar approach to finite size effects will
hold also in the non-abelian case briefly mentioned in [2].

The boundary conditions for the fermion fields in the path integral
have to be chosen with care; this was discussed at
length in ref [3] from a different point of view. Here we do not
wish to get into a general discussion of all possible different
boundary conditions. But, we shall see that the ``good'' choice we adopted
before, namely,
$$b^1_1=0;\ \ \ b^2_1=0;\ \ \ b^3_1=-{1\over 2};\ \ \ b^4_1=-{1\over 2};
\ \ \ \ \
b^1_2=0;\ \ \ b^2_2={1\over 2};\ \ \ b^3_2=0;\ \ \ b^4_2={1\over 2},
\eqno{(5)}$$
indeed is also ``good'' in that it leads to an answer compatible
with a clustering vacuum in the limit $l\to\infty$. However, the
calculation is done for arbitrary boundary conditions.

Our calculation proceeds similarly to one  
done for the vector Schwinger model,
presented in great detail in ref [4]. 
Since ours is a chiral case, we have to worry about phase
choices in the definition of various fermionic determinants,
and the generalization of [4] we need is not entirely 
trivial. To carry out the calculation, 
we have to resolve several ambiguities on the way.

Following [4], we decompose the gauge fields as
$$
A_1={\pi k\over l^2} x_2 + \partial_2 \phi + 
{2\pi\over l} h_1 + i g^{-1} \partial_1 g ;\ \ \ \ \ 
A_2= -{\pi k\over l^2} x_1 - \partial_1 \phi + 
{2\pi\over l} h_2 + i g^{-1} \partial_2 g \eqno{(6)}
$$
The terms dependent
on $g$ represent the gauge degree of freedom. $g$ is a periodic
function on the torus and takes values in $U(1)$. $g$ will disappear
from the calculation because of gauge invariance.
$\phi$ is periodic and has no zero momentum component - it describes the
non-uniform components of the electric field. 
The 
uniform components of the vector potential are represented by
the constants $h_\mu$. The uniform component of the electric
field is described by the $k$ dependent terms which are a
symmetric gauge representation of a configuration carrying flux $k$.
The space of gauge potentials falls into classes labeled by the integer
$k$. 

In [4] the various needed ``vectorial'' determinants were given 
definite values using $\zeta$-function regularization. For chiral
fermions we need to define complex square roots of these quantities.
At $k=0$ the required phase (a function of the $h_\mu$) is determined 
by the imposition of several symmetries [3,5,6]. At $k=1$ the fermionic
zero modes have to be separated out, 
and the remaining vectorial determinant
depends only on the volume of the system, which gives it the right
dimension. Now the phase freedom can be associated with the zero
modes and our choice will be explained below.

In what follows, it will be useful to define the function
$$\chiral(\alpha_1,\alpha_2;\tau) = 
\sum_{n=-\infty}^{\infty} \exp (-\pi\tau (n+\alpha_2)^2 + i2\pi n\alpha_1
+\pi i \alpha_1\alpha_2) \eqno{(7)} $$
$\chiral$ is closely related to the usual $\Theta$ function [4,5,6]. 
The chiral determinant and the zero modes 
are expressible in terms of $\chiral$.  

The expectation value of $V$ is the ratio of two
path integrals, one with $V$ inserted and the other with no
insertion. 
The path integral with no insertions
gets a contribution only from the 
$k=0$ sector and gives the partition function. After a change of
variables on the fermions and a computation of
the fermion determinant in the background of a
constant gauge potential [5,6], the partition function acquires the 
following form:
$$Z_0= {1\over \eta^5(1)} 
\int d^2h  \Biggl\{ \prod_f 
\chiral(h_1+b_1^f+{1\over 2}, h_2+b_2^f-{1\over 2}; 1) 
 \Biggr\}
  \Biggl\{
\chiral^*(2h_1+{1\over 2},2h_2-{1\over 2}; 1) 
\Biggr\} 
\int {\cal D}\phi e^{-\Gamma(\phi)};\eqno{(8)}$$
$$\Gamma(\phi)=
{1\over 2e_0^2} \int \phi (\Delta^2 - m_\gamma^2\Delta)\phi;
\ \ \ \ \ m_\gamma^2= {4e_0^2\over \pi};
\eqno{(9)}$$
$$
\eta(\tau)=
\exp (-{1\over 12} \pi\tau) \prod_{n=1}^\infty [1-\exp(-2\pi\tau n)].
\eqno{(10)}$$ 
The integral over $h_\mu$ represents a sum over saddles
in an otherwise Gaussian integral. 
The integral over $\phi$ contains the massive sector
of the theory and is identical to an integral that would appear in
a four flavor vector Schwinger model. 
The integrand in (8) is invariant under the
gauge transformation that takes $h_\mu\rightarrow h_\mu+1$ as long as
$$| \sum_f b_1^f | = | \sum_f b_2^f | = 1\eqno{(11)}$$ 
Clearly, (5) obeys this requirement. In [5] it was shown that
if anomalies cancel, invariance under $h_\mu\rightarrow h_\mu+1$
is guaranteed as long as all fermions obey antiperiodic boundary
conditions. Here this observation is slightly generalized
to other boundary conditions. It should be noted however that
not all boundary conditions are allowed. This restriction on boundary
conditions is distinct from the one discussed in [3].

To compute $\langle V\rangle$ we need to also compute the path integral
with the insertion of $V$. It gets a contribution only from the
$k=1$ sector. In this sector there are fermionic zero modes
and we need their explicit form. The zero modes at arbitrary
$\phi$ and $g$ are simply related to those presented below for
$\phi=0$ and $g\equiv 1$.

For
the $q=1$ fermions, the equation
$$\sigma_\mu (\partial_\mu + i A_\mu) \chi^0_f(x_1,x_2)=0\eqno{(12)}$$
has one solution. Normalizing, it takes the form
$$\chi^0_f(x_1,x_2)= {2^{1\over 4} \over l} 
e^ { \pi i \bigl[(b_1^f-h_1)(z_1-h_2)+ (b_2^f-h_2)(z_2+h_1) +{1\over 8}\bigr] }
\chiral(h_1+b_1^f+z_2, h_2+b_2^f - z_1; 1)
\eqno{(13)}$$
where $z_\mu={x_\mu\over l}$.
For the $q=2$ fermion, there are two linearly independent solutions to the equation
$$\sigma_\mu (\partial_\mu + 2 i A_\mu) \psi_p^0(x_1,x_2)=0;
\ \ \ \ p=0,1 \eqno{(14)}$$
which we pick (again normalized) of the form
$$\psi^0_p(x_1,x_2)= {\sqrt{2} \over l} 
e^ { \pi i \bigl[ 2h_1(h_2-z_1)+ (p-2h_2)(z_2+h_1) +{1\over 8} \bigr]}
\chiral(2h_1+2z_2, h_2-z_1+{p\over 2}; 2)
\eqno{(15)}$$
The above zero modes contain some phase choices we were free to make.
Since the phase choices contain an undetermined dependence on the $h_\mu$
what we choose has a nontrivial effect on the final answer. We recall
that the $h_\mu$ label different saddles. As such they play the role of
collective coordinates the zero modes depend on. From the vector case
we know that their role is to restore translational invariance
of the one point vertex we are computing. 
We extend this role also
to the phase choice.
The $h_\mu$ dependent phase factors are thus
added so as to make the solution a function of
$(z_1-h_2)$ and $(z_2+h_1)$. 
This leaves us with only one free constant which we pick so that   
the final answer corresponds to a vanishing $\theta$-parameter.
With all this in place, the path integral in the $k=1$ sector is
$$ \eqalign {Z_1= {
1\over l^4} & e^{-{2\pi^2\over e_0^2l^2}}\int d^2h 
     e^{ \pi i \bigl[(z_1-h_2)\sum_f b^f_1 + (z_2+h_1) \sum_f b^f_2 +{1\over 4}
\bigr] } \cr
 & \ \ \ \ \ \ \Biggl\{ \prod_f
\chiral( h_1+b_1^f+z_2 , h_2+b_2^f -z_1;1) 
  \Biggr\} \cr
 & 
\Biggl\{ \chiral^*(2h_1+2z_2,h_2-z_1;2) 
(\partial_1+i\partial_2)\Bigl[
e^{-\pi i (z_2 + h_1) }
\chiral^*(2h_1+2z_2,h_2-z_1+{1\over 2};2) \Bigr] \cr
&  - (\partial_1+i\partial_2) \Bigl[ \chiral^*(2h_1+2z_2,h_2-z_1;2)\Bigr]
e^{-\pi i (z_2 + h_1) }
\chiral^*(2h_1+2z_2,h_2-z_1+{1\over 2};2)
\Biggr\}
\cr
  \int {\cal D} & \phi e^{-\Gamma(\phi) -8\phi(x)}
\cr}
\eqno{(16)}
$$
where $z_\mu={x_\mu\over l}$ and $\partial_\mu$ means derivative with respect
to $z_\mu$.
$Z_1$  factorizes in a way similar to $Z_0$. 
The dependence on $m_\gamma$ comes in through
a factor which would be the same had we 
computed the 't Hooft vertex in the four flavor
Schwinger model.

Using an identity proven in the Appendix we can relate the $q=2$
contribution at $k=0$ with no insertion to the $q=2$ contribution
at $k=1$ in the presence of the insertion:
$$
\eqalign{
 & 
\Biggl\{ \chiral^*(2h_1+2z_2,h_2-z_1;2)
(\partial_1+i\partial_2 ) \bigl [
e^{-\pi i (z_2 + h_1) }
\chiral^*(2h_1+2z_2,h_2-z_1+{1\over 2};2) \bigr ] \cr
& \ \ \ \ \ \  -( \partial_1+i\partial_2 ) \bigl [
\chiral^*(2h_1+2z_2,h_2-z_1;2) \bigr ]
e^{-\pi i (z_2 + h_1) }
\chiral^*(2h_1+2z_2,h_2-z_1+{1\over 2};2)
\Biggr\} \cr
= & 4\pi \eta^3(1) e^{ \pi i (z_1-h_2-z_2-h_1-{1\over 4})}
\chiral^*(2h_1+2z_2-{1\over 2}, 2h_2-2z_1+{1\over 2}; 1) }
\eqno{(17)}
$$
We then obtain the expectation value of the 't Hooft vertex in our finite
volume:
$$\langle V\rangle_l = {64\pi \over (lm_\gamma)^4} 
{T_n\over T_d} \exp \Bigl[ -{4\pi\over lm_\gamma}
\coth \left ( {1\over 2}lm_\gamma \right ) \Bigr] 
e^{4F(lm_\gamma)-8H(lm_\gamma ,1)}.
\eqno{(18)}$$
The new $lm_\gamma$ dependent factors
come from the integration over $\phi$.
The functions 
$F(x)$ and $H(x,\tau)$ are given by 
$$\eqalign{
F(x)& =\sum_{n>0} \Bigl[ {1\over n} - {1\over \sqrt{n^2+(x/2\pi)^2}}\Bigr]\cr
H(x,\tau)& =\sum_{n>0}{1\over \sqrt{n^2+(x/2\pi)^2}}
{1\over e^{\tau\sqrt{(2\pi n)^2+x^2}}-1}.\cr}\eqno{(19)}$$
$T_n$ and $T_d$ are given by
$$\eqalign{ T_n = 
\int d^2h  & \Biggl\{ \prod_f 
\chiral(h_1+b_1^f+z_2, h_2+b_2^f-z_1; 1) 
 \Biggr\}
  \Biggl\{
\chiral^*(2h_1+2z_2-{1\over 2},2h_2-2z_1+{1\over 2}; 1) 
\Biggr\} \cr
& \ \ \ \ \
e^{\pi i \bigl[ (\sum_fb_1^f+1)(z_1-h_2) + (\sum_f b_2^f -1)(z_2+h_1)
\bigr]}
\cr}
\eqno{(20)}
$$
$$T_d = 
\int d^2h  \Biggl\{ \prod_f 
\chiral(h_1+b_1^f+{1\over 2}, h_2+b_2^f-{1\over 2}; 1) 
 \Biggr\}
  \Biggl\{
\chiral^*(2h_1+{1\over 2},2h_2-{1\over 2}; 1) 
\Biggr\}.
\eqno{(21)}
$$
Periodicity under $h_\mu\rightarrow h_\mu+1$ 
(which are gauge transformations)
of the integrand in $T_n$ is
assured by (11). 
On the other hand the integrand is only a function of the combinations 
$h^\prime_1=h_1+z_2$ and $h^\prime_2=h_2-z_1$, so it is actually
periodic with unit period in the $h^\prime_\mu$. Since
we integrate over a full fundamental domain the integral 
$T_n$ becomes independent of the 
$z_\mu$. We are then free to set $z_\mu={1\over 2}$
and observe that the integrand in $T_n$ becomes 
identical to the integrand in $T_d$ up to a phase.
The boundary conditions in (5) make this phase vanish
and imply $T_n = T_d$. 
Actually, (5) also implies that in both integrands one can
substitute for the $q=1$ factor (represented by the product
over ``flavors'' $f$) the complex conjugate of the
$q=2$ factor, rendering the integrands non-negative. 
The relevant identity was given in [3]. 

However, there are solutions to (11) which
differ from (5) and for which the phase
in the integrand of $T_n$ does not vanish. In that case one has to carry out
the integrals, which is not difficult, and one
gets a $T_n / T_d $ ratio different
from unity. Equation (22) below will change then and we shall
have no agreement with clustering at infinite volume. 
Note that these complications regarding boundary conditions
are absent in a vector theory. There flavor dependent boundary conditions
have no effect on the vertex. Still, the finite size 
{\it correction}, as a multiplicative factor, is universal
and can be obtained from the associated vector theory.

With our ``good'' boundary conditions, (5),
$T_n=T_d$ and (18) gives us the final answer for the 
vertex in a finite volume.
The infinite volume limit of (18) is
$$\langle V\rangle = {e^{4\gamma} \over 4 \pi^3} 
\eqno{(22)}$$
and agrees with the result in (4) quoted from [2] where it was
obtained asserting clustering.

Up to now we worked on a torus of equal sides. The formulae
can be generalized to the case of a torus of size $t\times l$
and the result is of potential use for future simulations. With
the boundary conditions of the same type as in (5) the
more general expression for the 't Hooft vertex on the torus
reads:

$$\langle V\rangle_{t\times l} = {64\pi \over (t m_\gamma)^4} 
\exp \Bigl[ -{4\pi\over  t m_\gamma}
\coth \left ( {1\over 2}lm_\gamma \right ) \Bigr] 
e^{4F( t m_\gamma)-8H( t m_\gamma , {t \over l} )}.
\eqno{(23)}$$

\noindent
{\bf 2.} Now that we have the finite volume 
result we can look at the data in [2]
closer since the only source of systematic error left is the
finiteness of the UV cutoff. The size of the lattice is $L$ in each 
direction
with $La=l$ where the lattice spacing is $a$. Our simulation was at
constant $l$ (in terms of the gauge coupling) and we attempted to 
take $a$ to zero
by letting $L$ grow. 

We discussed in [2] the possible appearance of a
Thirring term (of dimension 2)
and tuned a certain free parameter in the overlap to
make the induced Thirring coupling numerically negligible. 
So, we assume that there is no Thirring coupling and see if this assumption
is consistent with our data and the exact result. The remaining 
finite $a$ corrections come in integral powers of $a$. Such corrections
come from two sources: the operator and the action. The correction coming
from the operator have been dealt
with in [2] where we extracted a factor representing (quite sizable) UV
corrections due to the lattice point split representation of the 
derivative in $V$. Thus we are left to worry only about corrections
coming from the action. Exact global chiral symmetries are preserved
by the overlap and therefore only order $a^2$ corrections, coming from
dimension four operators are allowed.\footnote{*}{If one applies 
the overlap to the simulation of four dimensional
vector theories with massless quarks, one shall need no order $a$
improvement since chirality is exact. All errors are of order $a^2$.}
 
\figure{1}{\captionone}{5.0}

In figure 1 we show a plot with our data (with full gauge invariance implemented). We have data for 
$L=8,10,12,14,16$ and they are plotted against $1/L^2$ which is the same as $a^2$ for fixed $l$. To get a feel for how our numbers would extrapolate
to the continuum 
we computed a simple least square 
linear fit to the last four points and show the straight line so obtained.
The infinite $L$ limit is the continuum number computed
from (18) and is also shown on the graph. 

Our estimate, .0376 $\pm$ .0019, 
for the continuum limit value of the 't Hooft vertex in
the volume we worked at ($lm_\gamma = 3$)
comes from the intercept of the
linear fit. It  falls sufficiently close
to the exact result we get from (18), namely 0.0389, 
for us to be quite confident in our assumption about the
smallness of the effective Thirring coupling. Realistically,
we could have hardly expected a result more favorable to the overlap than
the one we obtained.

Therefore we would claim to have shown that the overlap quantitatively
reproduces fermion number violation in the 11112 model.

\vskip 1cm
\centerline{\bf Appendix}
\vskip .5cm
In this appendix we prove the identity written down in (17) which was
used in 
our calculation of the 't Hooft vertex.

Using the expression for the $\chiral$ function in (7) the LHS of
the (17) becomes
$$ \eqalign{
\sum_{n,m} [4\pi  (2m-2n+1)] & \ \ \ \exp 
\Bigl [ -2\pi (n^2 + m^2 ) - 4\pi (n+m)(h_2-z_1) - 2 \pi m \cr
& - 2 \pi (h_2-z_1)^2
-2\pi (h_2-z_1+{1\over 2})^2 - 4\pi i (n+m) (h_1+z_2) \cr
& -2\pi i (h_1+z_2)
(2h_2-2z_1+{1\over 2}) -\pi i (z_2+h_1)\Bigr]}\eqno{(A.1)}$$
We split the sum into cases.
When $(n+m)$ is even, we write $n=k+l$ and $m=k-l$. When $(n+m)$ is odd,
we write $n=k-l+1$ and $m=k+l$. As $n$ and $m$ range over all integers,
$k$ and $l$ also range over all integers. 
The LHS of (17) then becomes
$$\eqalign{
-
&\sum_l 4\pi  (4l-1) e^{-4\pi (l-{1\over 4})^2} \cr
& \sum_k e^{-\pi(2k+2h_2-2z_1+{1\over 2})^2 
-2 \pi i (2k)(2h_1+2z_2)
-\pi i(2h_2-2z_1+{1\over 2}) (2h_1+2z_2)
-\pi i (z_2+h_1)} \cr
+ 
&\sum_l 4\pi  (4l-1) e^{-4\pi (l-{1\over 4})^2} \cr
& \sum_k e^{-\pi(2k+1+2h_2-2z_1+{1\over 2})^2 
-2 \pi i (2k+1)(2h_1+2z_2)
-\pi i(2h_2-2z_1+{1\over 2}) (2h_1+2z_2)
-\pi i (z_2+h_1)} \cr
}\eqno{(A.2)}$$
where the first term corresponds to even $(n+m)$ and the
second term corresponds to odd $(n+m)$.
In both the terms in (A.2), the first factor is the same. The second factors
in both the terms can be combined into one sum since the first one have 
$2k$ and the second one has $2k+1$. Then we can write (A.2) as
$$\eqalign{ & 
\sum_l 16\pi  ({1\over 4}-l) e^{-4\pi (l-{1\over 4})^2} \cr
& \sum_n e^{-\pi(n+2h_2-2z_1+{1\over 2})^2 
-2 \pi i n(2h_1+2z_2-{1\over 2})
-\pi i(2h_2-2z_1+{1\over 2}) (2h_1+2z_2-{1\over 2})
+\pi i (z_1-h_2-z_2-h_1-{1\over 4})} \cr
=& 4\pi \eta^3(1) \cr
& \sum_n e^{-\pi(n+2h_2-2z_1+{1\over 2})^2 
-2 \pi i n(2h_1+2z_2-{1\over 2})
-\pi i(2h_2-2z_1+{1\over 2}) (2h_1+2z_2-{1\over 2})
+\pi i (z_1-h_2-z_2-h_1-{1\over 4})} \cr
}\eqno{(A.3)}
$$
(A.3) is obtained by employing the following identities 
$$\eta^3(1)  =  {1\over 2\pi} \theta^\prime_1(0,1) 
 = \sum_{n=-\infty}^\infty (-1)^n (n+{1\over 2}) e^{-\pi(n+{1\over 2})^2}
 = 4\sum_{l=-\infty}^{\infty} (l+{1\over 4}) e^{-4\pi (l+{1\over 4})^2}
\eqno{(A.4)}$$
The first equality in (A.4) is a well known classical identity.
For our purposes here, the second equality can be viewed simply as a
definition of the quantity  ${1\over 2\pi} \theta^\prime_1(0,1)$.
The last equality in (A.4) is obtained by splitting the sum over $n$ into
even $n=2l$ and odd $n=-(2l+1)$ where in both cases $l$ ranges from
$-\infty$ to $\infty$.
Using the definition of $\chiral$ from (7) in (A.3) results in the
RHS of (17).


\vskip .3in
\noindent {\bf Acknowledgments}:
The research of R. N. was supported in part by the 
DOE under grant \# DE-FG03-96ER40956 and \# DE-FG06-90ER40561 and that of
H. N. was supported in part by the DOE under grant \#
DE-FG05-96ER40559.

\if \preprint Y
\vskip .3in
\fi
\if \preprint N
\vskip 1in
\fi
\noindent{\bf References}
\vskip .2in
\item{[1]} D. Kutasov, A. Schwimmer, Nucl. Phys. B442 (1995) 447.
\item{[2]} R. Narayanan, H. Neuberger, hep-lat/9609031, Phys. Lett. B, to
appear.
\item{[3]} R. Narayanan, H. Neuberger, Nucl. Phys. B477 (1996) 521.
\item{[4]} I. Sachs, A. Wipf, Helv. Phys. Acta 65 (1992) 652.
\item{[5]} R. Narayanan, H. Neuberger, Phys. Lett. B348 (1995) 549.
\item{[6]} L. Alvarez-Gaume, G. Moore, C. Vafa, Comm. Math. Phys. 6 (1986) 1.
\item {} C. D. Fosco, S. Randjbar-Daemi, Phys. Lett. B354 (1995) 383.
                                                                      
\if \preprint N
\vskip 1in
\noindent{\bf Figure Captions}

\halign{#\hfill\qquad &\vtop{\parindent=0pt \hsize=5in \strut#
\strut}\cr
Figure 1: & \captionone\cr
&\cr}

\vskip .5in

\figureb{1}{7.0}

\fi
\vfill\eject
\end